\documentclass[aps,twoside]{revtex4}
\usepackage{epsfig}
\usepackage{amsmath,amssymb,color}
\usepackage[english]{babel}

\parskip=\medskipamount

\newcommand{\eq}[1]{(\ref{#1})}
\newcommand{\fig}[1]{Fig.\ref{#1}}
\newcommand{\be}{\begin{equation}}
\newcommand{\ee}{\end{equation}}
\newcommand\disp{\displaystyle}

\newcommand{\la}{\left<}
\newcommand{\ra}{\right>}

\def\runninghead#1#2{\pagestyle{myheadings}
\markboth{{\protect\it{\quad #1}}\hfill} {\hfill{\protect\it{#2\quad}}}}

\begin{document}

\runninghead{V.A. Avetisov, A.V. Chertovich, S.K. Nechaev, O.A. Vasilyev}{On scale--free and poly--scale
behaviors of random hierarchical network}

\title{On scale--free and poly--scale behaviors of random hierarchical networks}

\author{V.A. Avetisov$^1$, A.V. Chertovich$^2$, S.K. Nechaev$^3$\footnote{Also at: P.N.
Lebedev Physical Institute of the Russian Academy of Sciences, 119991, Moscow, Russia},
O.A. Vasilyev$^{4,5}$}

\affiliation{$^1$N.N. Semenov Institute of Chemical Physics of the Russian Academy of
Sciences, 1199911, Moscow, Russia \\ $^2$Physics Department, Moscow State University,
119992 Moscow, Russia \\ $^3$LPTMS, Universit\'e Paris Sud, 91405 Orsay Cedex, France \\
$^4$Max-Planck-Institut f{\"u}r Metallforschung, Heisenbergstr.~3, D-70569 Stuttgart,
Germany \\ $^5$Institut f{\"u}r Theoretische und Angewandte Physik, Universit{\"a}t
Stuttgart, Pfaffenwaldring 57, D-70569 Stuttgart, Germany}

\date{\today}

\begin{abstract}

In this paper the question about statistical properties of block--hierarchical random matrices is raised
for the first time in connection with structural characteristics of random hierarchical networks obtained
by mipmapping procedure. In particular, we compute numerically the spectral density of large random
adjacency matrices defined by a hierarchy of the Bernoulli distributions $\{q_1,q_2,...\}$ on matrix
elements, where  $q_{\gamma}$ depends on hierarchy level $\gamma$ as $q_{\gamma}=p^{-\mu \gamma}$
($\mu>0$). For the spectral density we clearly see the free--scale behavior. We show also that for the
Gaussian distributions on matrix elements with zero mean and variances $\sigma_{\gamma}=p^{-\nu \gamma}$,
the tail of the spectral density, $\rho_G(\lambda)$, behaves as $\rho_G(\lambda) \sim
|\lambda|^{-(2-\nu)/(1-\nu)}$ for $|\lambda|\to\infty$ and $0<\nu<1$, while for $\nu\ge 1$ the power--law
behavior is terminated. We also find that the vertex degree distribution of such hierarchical networks has
a  poly--scale fractal behavior extended to a very broad range of scales.

\bigskip


\end{abstract}

\maketitle

\section{Introduction}

The determination of the structural organization of the complex system with statistical disorder refers
often to a ``network paradigm'', implying the investigation of topological and statistical characteristics
of the contact map between the system elements. Analytical approaches to this task commonly use the
well--defined relationship between the spectral properties of the adjacency (connectivity) matrix, which
codes the contacts, and a particular topological structure of the network. In the case of uncorrelated
random networks (graphs) known as Erd\"os--R\'enyi (ER) graphs \cite{erdos}, there exists a bunch of
methods for analytical study of  spectra of the adjacency matrix -- from mathematically rigorous to less
rigorous. The last ones are often borrowed from the statistical mechanics of disordered systems. Among the
most used are the replica \cite{rogers} or the Flory mean--field \cite{paulk} approaches.

In the last decade, an explosive growth of computational power has allowed to accumulate a large body of
data on statistical and topological characteristics of real networks of diverse nature. The ``network
paradigm'' is currently spread from protein folding, intermolecular contacts in biopolymers, genetic maps
and cell metabolism, up to natural networks (including world wide web, various ecological, social,
financial and economic entities, etc.), statistical data analysis, and even to Bose--Einstein
condensation. Many of these topics are reviewed in \cite{barab}. In majority of cases it turned out that
the statistical characteristics of real network topology, as well as the spectral properties of the
adjacency matrices, essentially differ from the characteristics of random ER--graphs. For example, it has
been found that the probability distributions of typical topological characteristics  (e.g. the vertex
degree distribution, or the clustering coefficient) have the power--law tails for many real networks, in
contrast to the exponential tails for ER--graphs. Thus, the networks of such topology were stood out for a
class of the ``scale--free'' networks just because of the power--law behavior of some observables. In
practice, the network is often referred to belong to the ``scale--free'' class if the vertex degree
distribution (i.e. the distribution of nearest--neighboring contacts over the network vertices) has the
power--law tail. However in general setting dealing with the analysis of the spectral density of the
adjacency matrix we shall understand under the ``scale--free''--behavior the existence of the power--law
tail in the spectral density. The topological characteristics of these networks, such as, for example, the
vertex degree, have extremely wide distributions exceeding the corresponding distributions for ER--graphs
by orders of magnitude. Such anomalously wide distributions we shall call ``poly--scaled'' to distinct
them out of the ``scale--free'' behavior.

The main idea of our work consists in the construction of networks with block--hierarchical adjacency
matrices resulting from an appropriate randomization of the standard Parisi matrix -- one of the key
objects in the theory of spin glasses (see \cite{mezard} for example). In other words, we study below the
random block--hierarchical (RBH) networks. By the behavior of the spectral density of adjacency matrices,
the RBH networks fall into the ``scale--free'' class, but in majority of cases, they have ``poly--scaled''
distribution of the vertex degree.

It would not be an exaggeration to say that the block--hierarchical ``ordering'' is rather typical than
exceptional for many complex systems which are both random and multi--scaled. The examples of such
networks can be easily find in different areas of mathematics, physics and biology: from chaotic maps in
Hamiltonian systems \cite{chirikov,meiss} to condensed (globular) structures of polyelectrolyte chains
\cite{dobr} and hierarchical organization of biopolymers \cite{hierar_bio}. We anticipate our
consideration of spectral properties of the RBH adjacency matrices with a generic example of possible
realization of a block--hierarchical contact map in a globular phase of unknotted ring polymer molecules
with topological interactions.

It is known that the non-phantomness of a polymer chain causes two types of interactions: i) volume
interactions vanishing for infinitely thin chains, and ii) topological interactions, which present even
for chains of zero thickness. For sufficiently high temperatures, a polymer molecule strongly fluctuates
without reliable thermodynamic state called a coil state. However for temperatures below some critical
value, $\theta$, the polymeric chain exhibits a dense weakly fluctuating globular (drop--like)  structure
\cite{lif,lgkh}. In classical works \cite{lif,lgkh} devoted to the  coil-to-globule phase transition
without topological constraints, it has been shown that, for $T<\theta$, the globular state can be
described by using just only two-- and three--body interaction constants: $B=b\frac{T-\theta} {\theta}<0$
and $C={\rm const}>0$ (see \cite{grosbk,cloiz}). The approach developed in \cite{lif,lgkh} is regarded as
the basic one in modern statistical theory of collapsed polymeric state.

\begin{figure}[ht]
\epsfig{file=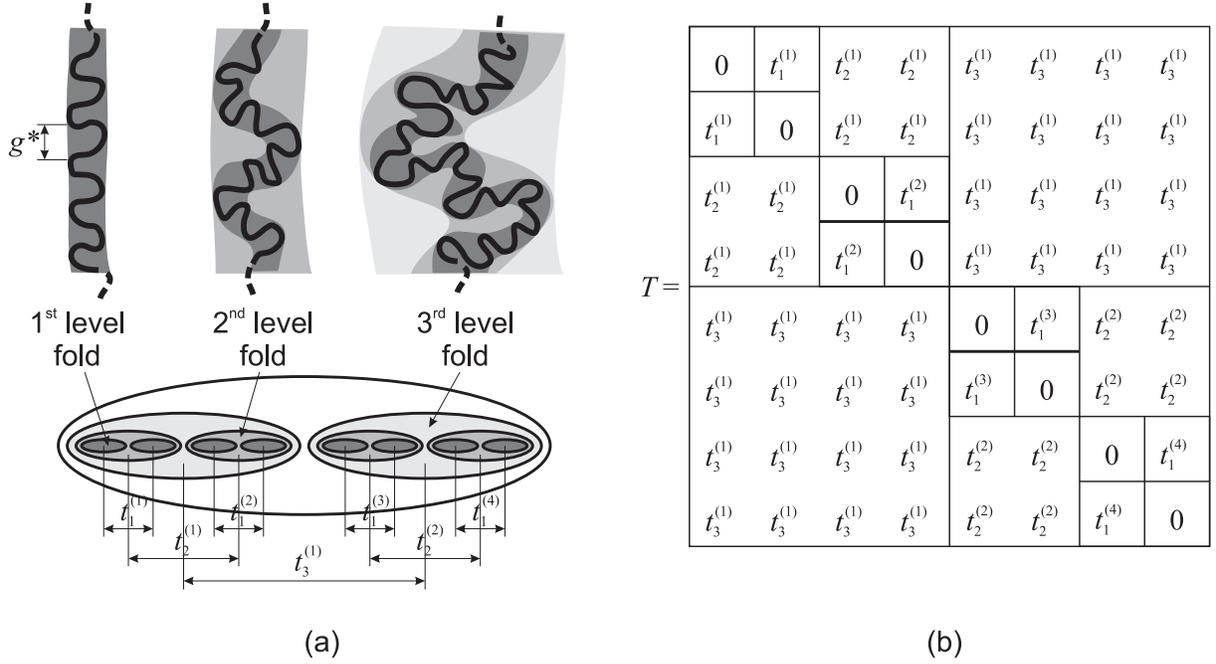,width=16cm} \caption{(a) Three subsequent stages of the construction of
hierarchical contact map for the "crumpled globule" (see the text for details); (b) Block--hierarchical
$p$--adic Parisi matrix $T$ ($p=2$).}
\label{parisi}
\end{figure}

The topological constraints in the globular phase of an unknotted macromolecule act the part of repulsive
interactions. For the temperatures below $\theta$--point (i.e. in a poor solvent), against the temperature
and the energy of volume interactions there exists certain scale of the chain length, $g^*$,  such that
the chains longer than $g^*$ collapse. Taking an enough long  chain, we can define these $g^*$--unit parts
as new ``block monomers'' (or ``folds of minimal scale''). In the \fig{parisi}a they are denoted as  the
1st level folds. Sufficiently long parts of the chain  with several folds of minimal scale  should again
``collapse in itself'', i.e. they should form the 2nd level folds, if other chain parts do not interfere
with it. The chain of such new sub-blocks of the 2nd level folds collapses again forming the 3rd level
folds, and so on... This block-hierarchical folding is completed when the initial chain units are united
into one fold of the largest scale. Three first consecutive steps of such a process are shown in
\fig{parisi}a. Note that the line representing the chain folded by this way resembles the 3D--analogue of
the well known self--similar {\it Peano curve}. The specific feature of the crumpled globule consists in
the fact that different chain parts are not entangled with each others, completely fill the allowed volume
of space and are "collapsed in themselves" starting from the characteristic scale $g^*$.

The scale of the fold, shown by hues of gray in \fig{parisi}a, can be considered as a cutoff for the
interaction distance between $g^*$--block monomers in the current fold. The values $t_{\gamma}^{(n)}$  can
be thought as interaction constants between these $g^*$--block monomers in the $\gamma$--level fold and,
thus, they constitute the contact map $T$ depicted in \fig{parisi}b.

In \cite{gns} and later, more rigorously, in \cite{nech_vas} it has been argued that the absence of knots
in a densely packed polymer ring causes a very peculiar fractal structure of the chain trajectory,
strongly affecting all thermodynamic properties of the macromolecule in the globular phase. The
corresponding structure of a collapsed unknotted polymer ring was called a {\em crumpled} globule. The
chain trajectory in the crumpled globule densely fills the volume such that all part of the chain become
segregated from each other in a broad region of scales. This model has been used later (see, for instance,
\cite{gr_rab,chromatin}) to describe self--similar hierarchical organization in some biopolymers, like DNA
and chromatin.

Below we discuss some topological properties of block--hierarchical random networks. On the basis of
obtained results we propose the new way of building of hierarchical networks with scale--free and
poly--scaled properties in a memoryless locally uniform way. This method does not demand the control of
the current state of the network.

\section{``Heavy tails'' in the spectral density of random block--hierarchical adjacency matrices}

Let us start with a description of generic procedure of the RBH--network construction. Taking $N$ points
as potential vertices of our forthcoming network, we raise a hierarchical network by connecting the
vertices by edges in a specific way. In the outset, we introduce an ensemble of the $N\times N$ adjacency
matrices, $T$; each of them encoding the edges between connected vertices in a network realization.
Namely, the element $T_{i,j}$ of $T$ is 1 if the vertices $i$ and $j$ are directly connected, otherwise
$T_{i,j}=0$. We consider the adjacency matrix in very peculiar form of a $p$--adic
translation--noninvariant Parisi matrix. This matrix is shown in \fig{parisi}b for $p=2$. Obviously,
$T_{i,j}=T_{j,i}$ and $T_{i,i}=0$. All matrix elements, $t_{\gamma}^{(n)}$, are the Bernoulli distributed
random variables:
\be
t_{\gamma}^{(n)} = \begin{cases} 1 & \mbox{with the probability $q_{\gamma}$} \\
0 & \mbox{with the probability $1-q_{\gamma}$} \end{cases}
\label{eq:1}
\ee
where $\gamma$ counts the hierarchy levels ($1\le \gamma \le \gamma_{\rm max}\equiv \Gamma$) and $n$
enumerates different blocks corresponding to a given hierarchy level $\gamma$ (see \fig{parisi}b). Note
that the probability $q_{\gamma}$ does not depend on $n$. The full ensemble of $N\times N$ matrices $T$,
where $N=p^{\Gamma}$, is completely determined by the set of probabilities, $\{Q\}=\{q_1, q_2,...,
q_{\Gamma}\}$. Thus, the elements $T_{i,j}$, being the random variables, are hierarchically organized {\em
in probabilities}. Below we consider the set of probabilities, $\{Q\}$, with $q_{\gamma}=p^{-\mu \gamma}$
($\mu>0$).

The systematic study of statistical properties of ensembles of random graphs (networks) deals with the
investigation of the spectral properties of a graph adjacency matrix \cite{fark,goh}. Let $\lambda_i$
($1\le i \le N$) be the eigenvalue of the adjacency matrix. The spectral density of the ensemble of random
symmetric adjacency matrices is defined in the standard way,
\be
\rho(\lambda)=\frac{1}{N}\sum_{i=1}^N \la \delta(\lambda-\lambda_i)\ra_{\{q_1, q_2,..., q_{\Gamma}\}}
\label{eq:2}
\ee
where $\la...\ra_{\{q_1,q_2,...q_n \}}$ denotes the averaging over the distributions of the matrix
elements, $t_{\gamma}^{(n)}$.

\begin{figure}[ht]
\epsfig{file=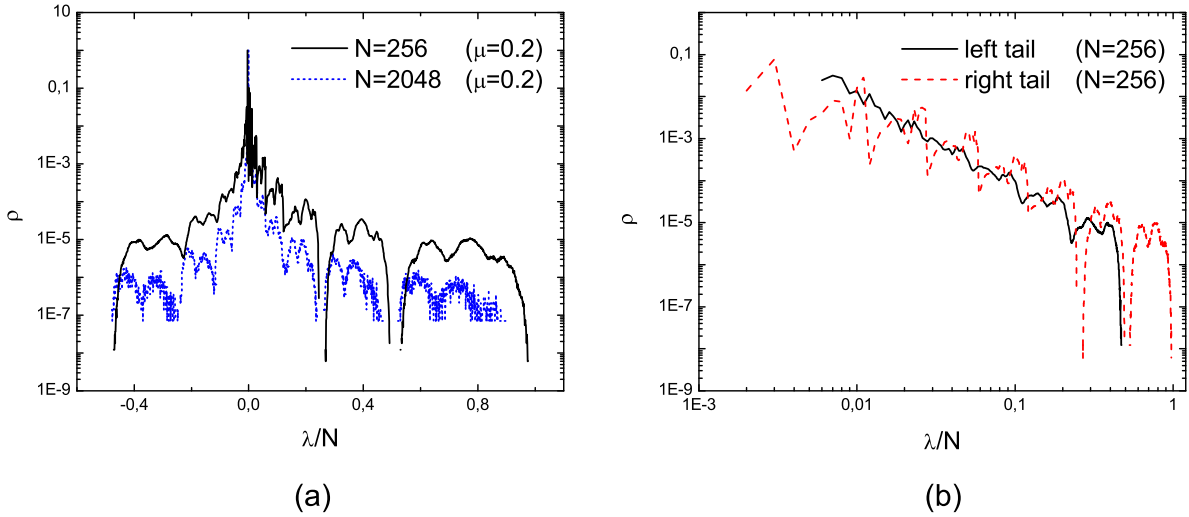,width=16cm} \caption{Spectral density $\rho(\lambda)$ for $\mu=0.2$: (a) semi--log
plot of the full distribution for $N=256$ (solid line) and $N=2048$ (dashed line), (b) the left-- and
right--hand tails of $\rho(\lambda)$ for $N=256$ in log--log coordinates.} \label{fig:3}
\end{figure}

Computing numerically the spectral density, $\rho(\lambda)$, of networks with block--hierarchical
adjacency matrices, we found that the tails of the spectral density $\rho(\lambda)$ follow a power--law
asymptotic behavior $\rho(\lambda)\sim |\lambda|^{-\chi}$ with the exponent $\chi=\chi(\mu)$. The sample
plots of the spectral density $\rho(\lambda)$ are shown for $N=256, 2048$ and $\mu=0.2$ in \fig{fig:3}a in
semi--log coordinates. The corresponding log--log plot of the left-- and right--hand tails of the spectral
density for $N=256$ is drawn in \fig{fig:3}b. It is interesting to note that the right--hand tail of
$\rho(\lambda)$, while demonstrating the same behavior, is worse averaging.

Some analytic arguments supporting the found power--law behavior of the spectral statistics of the RBH--
graphs can be brought by means of the spectral density, $\rho_G(\lambda)$, of the Gaussian ensemble of the
Parisi matrices $T$, where the Bernoulli distributions on matrix elements of $T$ are replaced by the
Gaussian distributions with zero means and a set of variances $\{{\cal S}\}=\{\sigma_1, \sigma_2,...,
\sigma_{\Gamma}\}$,  $\sigma_{\gamma}=p^{-\nu\gamma}$ ($\nu>0$). To get the spectral density
$\rho_G(\lambda)$, note that the eigenvalues of the standard (i.e. translation--invariant) Parisi matrix
with $T_{i,i}=0$ can be expressed in terms of matrix elements $t_{\gamma}^{(n)}\equiv t_{\gamma}$ as
follows (\cite{ogi}):
\be
\lambda_{\gamma}=p^{\gamma} t_{\gamma}-(1-p^{-1}) \sum_{\gamma'=1}^{\gamma} p^{\gamma'} t_{\gamma'} \qquad
(\gamma=1,..., \Gamma)
\label{eq:3}
\ee
The eigenvalue $\lambda_{\gamma}$ is $p^{\Gamma-\gamma}$ times degenerated ($\gamma=1,...,\Gamma$). In
addition there is one extra eigenvalue $\lambda_0=-\sum_{\gamma=1}^{\Gamma} p^{\Gamma-\gamma}
\lambda_{\gamma}$. For the translation--noninvariant Parisi matrix we generalize \eq{eq:3} in the way
similar to the one used for  block--hierarchical kinetic matrices (see \cite{avetisov}). Remind that for
eigenvalues of {\em kinetic} matrix, one has
\be
\lambda_{\gamma,n}=-p^{\gamma} t_{\gamma}^{(n)} - (1-p^{-1}) \underbrace{\sum_{\gamma'=\gamma+1}^{\Gamma}
p^{\gamma'} t_{\gamma'}^{(n')}}_{\Sigma}
\label{eigen}
\ee
i.e. the eigenvalue $\lambda_{\gamma,n}$ can be expressed via a linear combination ($\Sigma$ in
\eq{eigen}) of weighted matrix elements $t_{\gamma'}^{(n')}$ coming from the vertices $(\gamma',n')$ along
a unique path on the $p$--adic Cayley tree from the vertex $(\gamma'=\gamma, n'=n)$ towards the root
vertex $(\gamma'=\Gamma, n'=1)$ -- see \cite{nech} for more details. In the framework of same
geometrical interpretation, the eigenvalue $\lambda_{\gamma, n}$ of the block--hierarchical matrix with
$T_{i,i}=0$ (shown in \fig{parisi}b) seems to be the linear combination of weighted matrix elements
$t_{\gamma'}^{(n')}$ along a path on the $p$--adic Cayley tree from the bottom level $\gamma'=1$ to the
vertex $(\gamma'=\gamma, n'=n)$. However this construction has an ambiguity since the corresponding path
is not uniquely defined by the pair $(\gamma,n)$. So, rigorously speaking, the eigenvalues of particular
realization of the random block--hierarchical adjacency matrix cannot be parameterized by the pairs
$(\gamma,n)$, and we cannot write an exact expression for the eigenvalues in the form of expression
\eq{eq:3}. However  for  computation of  spectral density, $\rho_G(\lambda)$, we use below {\em a
posteriori} self--averaging arguments, which make our consideration self--consistent. Moreover, the
extensive numerical simulations confirm our analytic prediction of $\rho_G(\lambda)$ for $|\lambda|\gg 1$
in the interval $0<\nu<1$. Thus, formally extending \eq{eq:3} to the case of $\lambda_{\gamma,n}$ of
block--hierarchical matrix $T$, we replace in \eq{eq:3} the first term by $t_{\gamma}^{(n)}$ and the sum
-- by $\sum_{\gamma'=1}^{\gamma} p^{\gamma'} t_{\gamma}^{(n')}$ where the summation runs now along the
paths on the Cayley tree from the hierarchical level $\gamma'=1$ to the vertex $(\gamma,n)$ located on the
hierarchical level $\gamma'=\gamma$ (compare to \eq{eigen}). Supposing the distribution of the matrix
elements $t_{\gamma}^{(n)}$ to be Gaussian,
\be
P(t_{\gamma}^{(n)})=\frac{1}{\sqrt{\pi \sigma_{\gamma}^2}}
\exp\left(-\frac{(t_{\gamma}^{(n)})^2}{\sigma_{\gamma}^2}\right)
\label{eq:5}
\ee
and using for $\lambda_{\gamma,n}$ the guessed expression as a linear combination of matrix elements, we
end up with the following equation for the spectral density, $\rho_G(\lambda)$:
\be
\rho_G(\lambda)=p^{-\Gamma}\sum_{\gamma,n}\la \delta(\lambda-\lambda_{\gamma,n})
\ra_{P(t_{\gamma}^{(n)})}=\frac{1}{\sqrt{\pi}} \sum_{\gamma=1}^{\Gamma} p^{-\gamma}
\frac{1}{\sqrt{u_{\gamma}^2}} \exp\left(-\frac{\lambda^2}{u_{\gamma}^2}\right)
\label{eq:4}
\ee
where for $\sigma_{\gamma}=p^{-\nu \gamma}$ we have:
\be
u_{\gamma}^2= p^{2(\gamma-1)}\sigma_{\gamma}^2 + (1-p^{-1})\sum_{\gamma'=1}^{\gamma-1} p^{2\gamma'}
\sigma_{\gamma'}^2= \frac{p-2}{p} p^{2(1-\nu)\gamma}+ \frac{(p-1)^2}{p-p^{\nu}} (p^{2(1-\nu)\gamma}-1)
\label{eq:7}
\ee
For $p=2$ we can rewrite $\rho_G(\lambda)$ in \eq{eq:4} for $\Gamma\to \infty$ as follows
\be
\rho_G(\lambda)\simeq \frac{1}{\sqrt{\pi}} \sum_{\gamma=1}^{\infty} 2^{-(2-\nu)\gamma}
\exp\left[-\lambda^2\frac{4-4^{\nu}} {2^{2(1-\nu)\gamma}-1}\right]
\label{eq:8}
\ee
Taking into account that
\be
\sum_{\gamma=1}^{\infty} p^{-c_1 \gamma}\, e^{-t p^{-c_2 \gamma}} \simeq t^{-c_1/c_2} \qquad (t\gg 1)
\label{eq:9}
\ee
and substituting $c_1=2-\nu$ and $c_2=1-\nu$, we arrive at the following asymptotic form for the spectral
density at $|\lambda|\gg 1$:
\be
\rho_G(\lambda)\simeq |\lambda|^{-\xi(\nu)} \qquad (0<\nu<1)
\label{eq:10}
\ee
where
\be
\xi(\nu)=\frac{2-\nu}{1-\nu}
\label{eq:10a}
\ee

\begin{figure}[ht]
\epsfig{file=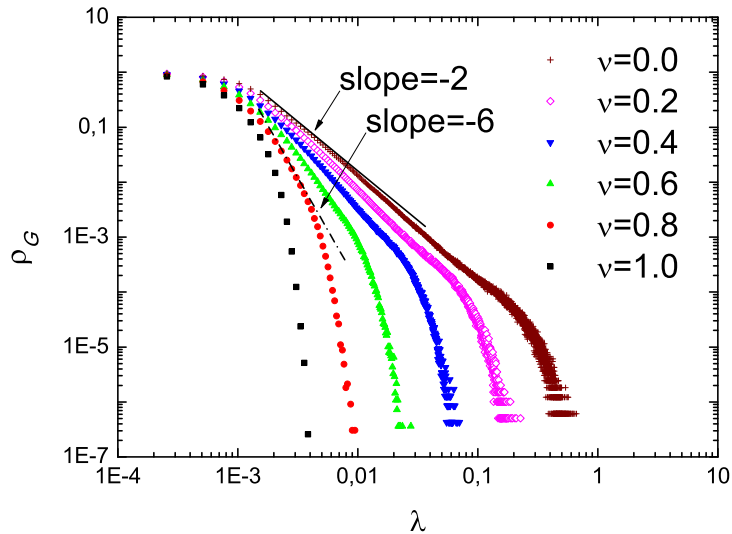,width=10cm} \caption{Spectral density for symmetric Gaussian translationary
noninvariant block--hierarchical matrices. The solid and dot--dashed lines have the slopes $\xi(\nu=0)=2$
and $\xi(\nu=0.8)=6$; scatter graphs: $\nu=0$ ($+$), $\nu=0.2$ ($\lozenge$), $\nu=0.4$
($\blacktriangledown$), $\nu=0.6$ ($\blacktriangle$), $\nu=0.8$ ($\bullet$), $\nu=1.0$ ($\blacksquare$).}
\label{fig:5}
\end{figure}

The arguments supporting our derivation of the expression \eq{eq:7} for the spectral density
$\rho_G(\lambda)$ are as follows. First of all, note that \eq{eq:4}--\eq{eq:7} become exact if we skip the
dependence on $n$ in the matrix elements $t_{\gamma}^{(n)}$ and, hence, restore the translational
invariance in the block--hierarchical matrix $T$. Secondly, we found in the extensive numeric simulations
summarized in \fig{fig:5} that indeed the conjectured behavior \eq{eq:10}--\eq{eq:10a} actually holds for
translation noninvariant Parisi matrices. In \fig{fig:5} we have plotted the tails of the spectral density
$\rho_G(\lambda)$ for Gaussian ensemble of block--hierarchical matrices for $N=256$. The solid and
dot--dashed lines have the slopes $-\xi(\nu)$, where $\xi(\nu=0)=2$ (solid line) and $\xi(\nu=0.8)=6$
(dashed line). The scatter graphs from top to bottom correspond to $\nu=0.0, 0.2, 0.4, 0.6, 0.8, 1.0$.

The fact that \eq{eq:10} gives right asymptotic behavior occurs apparently due to an effective
self--averaging of the sum of matrix elements along each particular path on a Cayley tree for the
distribution $\{{\cal S}\}=\{\sigma_1,\sigma_2,...\}$. One sees that in the sum in \eq{eq:7} for
$\sigma_{\gamma}=p^{-\nu \gamma}$ the lower limit of the summation can be shifted from $\gamma'=1$ to
$\gamma'\to -\infty$. Asymptotically the result for $u_{\gamma}^2$ will remain unchanged if $0<\nu<1$.
Such an extension of summation means that the computation of the spectral density (and, in particular, of
$\lambda_{\gamma,n}$) involves the summation along the infinite paths running from $-\infty$ to the
hierarchical level $\gamma$. We expect that for $\sigma_{\gamma}=p^{-\nu \gamma}$ ($0<\nu<1$) due to the
convergence of the sum $\sum_{\gamma'=-\infty}^{\gamma} p^{2\gamma'} \sigma_{\gamma'}^2$, the eigenvalue
$\lambda_{\gamma,n}$ does not depend on each particular path on a Cayley tree and hence, the eigenvalue
does not depend on the index $n$. Once this point of view is accepted, we return to an effective
translation invariant block--hierarchical matrix for which \eq{eq:8} is exact.

One can see from \eq{eq:10}--\eq{eq:10a} and \fig{fig:5} that for $\nu\ge 1$ the power--law behavior of
the spectral density $\rho_G(\lambda)$ terminates. This termination deserves special attention. Indeed,
for $\nu\ge 1$ we cannot extend  the lower limit of summation over $\gamma'$ to $-\infty$ since the
corresponding sum diverges. Hence, the contribution to the eigenvalues (and, therefore, to the spectral
density) strongly depend on the particular configuration of the path. In this case the self--averaging,
suggested in our computation, is invalid anymore and we cannot say anything about the behavior of the
spectral density in the region $\nu\ge 1$. The opposite case $\nu<0$ deserves special attention because
formally the sum \eq{eq:7} converges for any $\nu<0$ and our arguments about self--averaging seem to work.
However the details of the analysis of this case is beyond the scope of the current work and will be
discussed elsewhere.

Our approach to the construction of the networks with the scale--free behavior in the spectral density is
not unique. As it is shown in  the recent work \cite{bohigas}, the procedure of dividing Gaussian matrices
by a random variable, as well as the same procedure applied to random graphs, leads to the spectral
density interpolating between the Erd\"os-R\'enyi and the scale--free models.

\section{Hierarchical ``growing'' of poly--scaled networks}

One can roughly distinguish two methods of construction the scale--free networks. The first method has
been developed mainly for illustrative purposes and deals with the hierarchical construction \cite{rav} of
deterministic scale--free graphs with predetermined fractal properties such as, say, vertex degree
distribution. Since in this case the obtained graphs are deterministic, it is senseless to talk about any
statistics of their spectra. The second method deals with the variants of iterative ``preferential
attachment'' construction \cite{pref}, where new nodes are added to a vertex of the network with the
probability depending on already existing vertex degree (the number of one--step connections to other
vertices). Almost all known statistical characteristics of scale--free networks, including the spectral
density of adjacency matrix, are obtained for the networks constructed using this method. Typically, the
spectral density of the ensemble of scale--free networks designed by the preferential attachment method
has a triangle--like shape in the "bulk" part with power--law tails.

Note that the construction of scale--free networks by preferential attachment method is based on locally
nonuniform incremental growth with unlimited evolutionary memory. In contrast to this, we propose below
another physically motivated approach to the construction of the random networks with power--law spectral
density by a ``parallel'' (i.e. non-stepwise) and uniform procedure.

The peculiarity of our construction consists in the following. We build clusters of {\em edges} with
hierarchically organized probabilities, while the typical procedure consists in hierarchical grouping of
{\em vertices}. Our procedure does not impose any additional metric structure on the graph and leaves the
graph (network) purely topological. To the contrary, grouping of vertices imposes a metric structure on
the graph (network) since such a grouping operates usually with the notion of ``close'' (or ``distant'')
vertices.

To shed light on spectral properties of random hierarchical graphs we exploit the link between the Random
Matrix Theory (RMT) and the Graph Theory (GT). It is known that the spectral density $\rho(\lambda)=
\frac{1}{N}\sum_{i=1}^N \la \delta(\lambda-\lambda_i)\ra_{\{q_1, q_2,..., q_{\Gamma}\}}$ of the ensemble
of adjacency matrices is directly related to the topological structure of the corresponding network since
the value
\be
M_k=\frac{1}{N}\int \lambda^k \rho(\lambda) d\lambda=\frac{1}{N}\sum_{i=1}^N \lambda_i^k
\label{eq:2a}
\ee
defines (up to the factor $N$) the average number of $k$--step loops in the network (see, for example,
\cite{fark}).

\begin{figure}[ht]
\epsfig{file=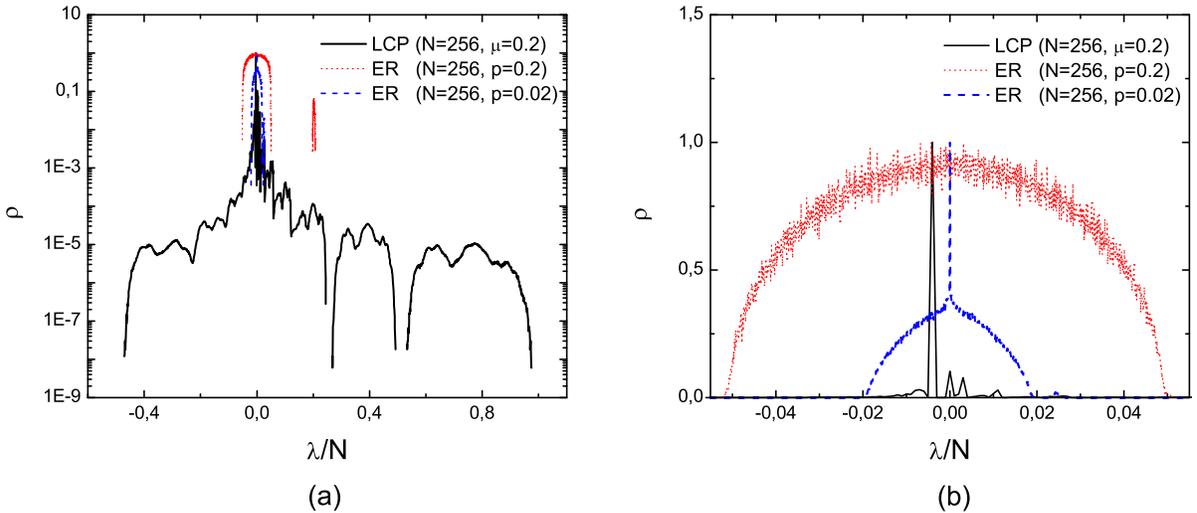,width=16cm} \caption{Comparison of spectral densities of random
block--hierarchical matrix and Erd\"os--R\'enyi (ER) random graphs: (a) in semi--logarithmic coordinates
in the region $-0.5<\lambda/N<0.5$; and (b) in linear coordinates in the region $-0.05<\lambda/N<0.05$.}
\label{fig:4a}
\end{figure}

In particular, in \cite{komlos} it has been shown that in the thermodynamic limit the spectral properties
of random Erd\"os--R\'enyi graphs \cite{er} coincide with the spectral properties of random real symmetric
matrices. This result is one of the benchmarks in our consideration. The elements of adjacency matrix,
$A_{i,j}$, of Erd\"os--R\'enyi graph are Bernoulli--distributed random variables: $A_{i,j} = 1$ or 0 with
probabilities $q$ and $ 1-q$ correspondingly. For the ensemble of ER--random graphs the density,
$\rho_A(\lambda)$, of eigenvalues of adjacency matrices $A$ can be analytically computed in the
thermodynamic limit $N\to\infty$ and has for some $q$ a celebrated Wigner--Dyson semicircle law known for
ensembles of Gaussian matrices \cite{wigner}. Namely, the following statement is proved \cite{komlos}. Let
$B$ be a real symmetric $N\times N$ matrix with independently distributed entries $B_{i,j}$ from, say,
Gaussian distribution $P(B_{i,j})$ with $\la B_{i,j} \ra =0$ and $\la B_{i,j} \ra = \sigma^2$. Then the
spectral density, $\rho_B(\lambda)$, of the ensemble of matrices $B$ converges in the limit $N\to\infty$
to the semicircle distribution
\be
\rho_B(\lambda)=\begin{cases} \disp \frac{1}{2\pi \sigma^2} \sqrt{4 N \sigma^2-\lambda^2} & \mbox{if
$|\lambda|<\sqrt{4 N\sigma^2}$} \medskip \\ 0 & \mbox{if $|\lambda|>\sqrt{4 N\sigma^2}$}
\end{cases}
\label{semicirc}
\ee
If $\sigma^2=q(1-q)$, then $\rho_A(\lambda)=\rho_B(\lambda)$ for $N\to\infty$, i.e. the spectral densities
of ensembles of random ER--graphs and of Gaussian symmetric matrices coincide in the thermodynamic limit.
Nevertheless, such a coincidence of spectral densities for random graphs and random Gaussian matrices
should not be understood in a literal sense: some spectral properties of random ER--graphs and random
matrices are different \cite{fark,many}. For example, since for the adjacency matrix $A$ of random
ER--graphs one has $\la A_{i,j} \ra =q$, then the corresponding largest eigenvalue, $\lambda_1$, grows
linearly with the system size, $N$, i.e. $\lambda_1 = Nq$, meaning that the semicircular distribution for
random graphs is valid only for the matrix $A-\la A \ra$. Also, the tails of spectral distributions near
the spectrum edges are different for random graphs and random matrices. Nevertheless, as a first
approximation, the random Gaussian matrices could serve as a very natural benchmark for the corresponding
statistical analysis. The comparison of spectral properties of hierarchical and random Erd\"os--R\'enyi
graphs is demonstrated in \fig{fig:4a}. The \fig{fig:4a}a shows the semi--log plot of the spectral
densities for: (i) random hierarchical graphs with LCP adjacency matrices for $N=256$ and $\mu=0.2$ (solid
line), and (ii) random Erd\"os--R\'enyi graphs for $N=256$ and $p=0.2$ (dotted line), for $N=256$ and
$p=0.02$ (dashed line). In \fig{fig:4a}b we have redrawn the central part of \fig{fig:4a}a in the linear
scale.

\begin{figure}[ht]
\epsfig{file=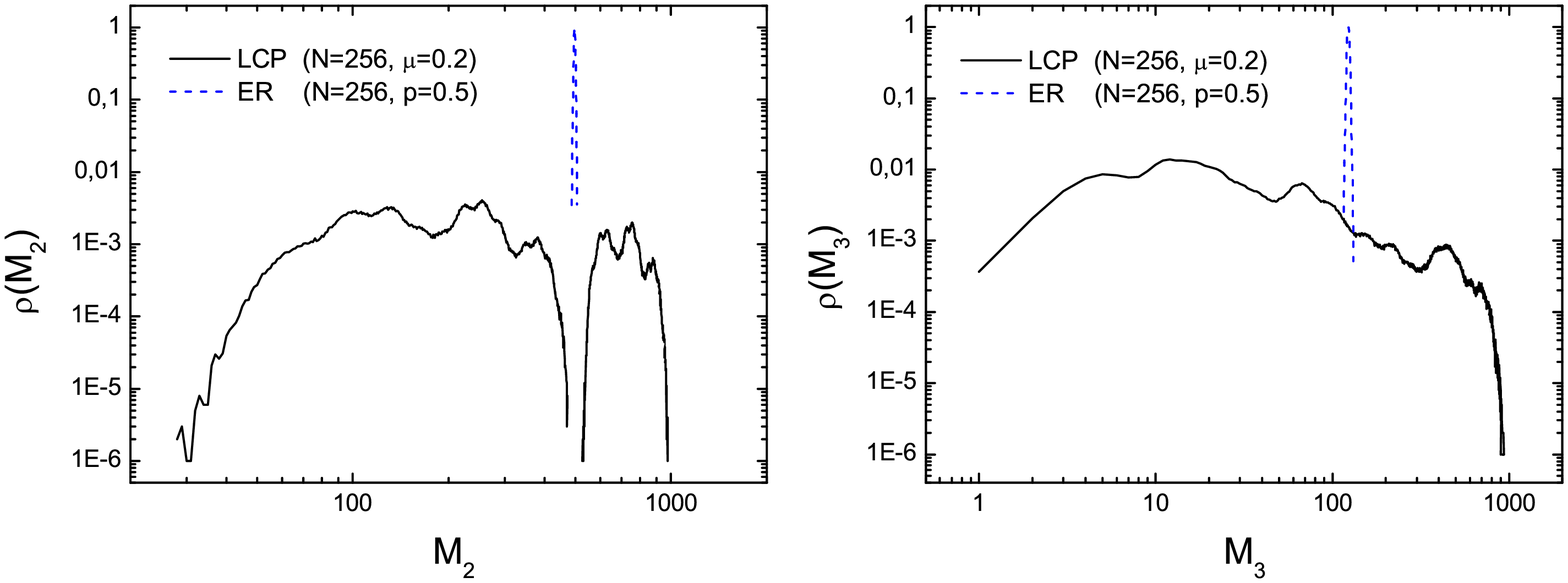,width=16cm} \caption{Distribution of the number of 2-- and 3--step loops in random
hierarchical and Erd\"os--R\'enyi graphs.}
\label{fig:4}
\end{figure}

The numerical results for the probability distributions $P(M_k)$ ($k=2,3$) of the number of $k$--step
loops in random hierarchical graphs for $N=256$ and $\mu=0.2$ are compared in \fig{fig:4} with the
distributions of loops on Erd\"os--R\'enyi random graphs for $N=256$ and $p=0.2$. Recall that $P(M_2)$
defines the probability to have in the finite graph the average connection degree equal to $M_2$.

One sees that the distribution functions $P(M_2)$ and $P(M_3)$ for our hierarchical random graphs are much
broader than the corresponding distributions for Erd\"os--R\'enyi graphs with the same number of vertices.
Hence, the topological structure of random hierarchical graphs is much more "flexible" than that of random
ER--graphs. This is consistent with the found behavior for the spectral density: the distribution function
$\rho(\lambda)$ has ``heavy'' tails and decays much slower than the that for random ER--graphs. According
to the behavior of the distribution functions $P(M_2)$ and $P(M_3)$ it is naturally to call our random
block--hierarchical graph the ``poly--scaled''.

\begin{figure}[ht]
\epsfig{file=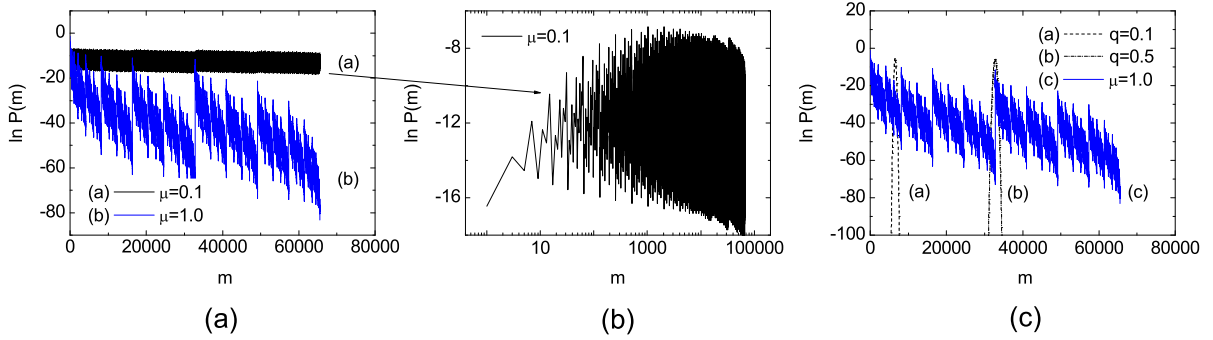,width=16cm} \caption{a) The family of distributions ${\cal P}(m)$ for $\Gamma=16$
at $\mu=0.1; 1.0$; b) Distribution ${\cal P}(m)$ in the double--logarithmic scale for $\mu=0.1$ and
$\Gamma=16$; c) Comparison of ${\cal P}(m)$ for the random block--hierarchical graph ($\mu=1.0$,
$\Gamma=16$) and ${\cal P}_{\rm ER}(m)$ for the Erd\"os--R\'enyi graph ($q=0.1; 0.5$, $N=2^{16}$).}
\label{2:vdd_fig}
\end{figure}

The method of generating functions allows us to compute easily the vertex degree distribution in the
ensemble of random block--hierarchical graphs directly by their adjacency matrices. To do that let us
consider any (for example, the first) row in the adjacency matrix $T$ (see \fig{parisi}b). The total
number of links of the first graph vertex to other vertices is defined by the number of matrix elements,
$t_{\gamma}^{(1)}$ ($\gamma=1,...,\Gamma$), having in the first row nonzero value (i.e. taking the
value ``1''). Thus, the distribution of the number of connections (i.e. vertex degree distribution),
${\cal P}(m)$, is the probability of the fact that the sum of matrix elements in the first row is exactly
equal to $m$ under the condition that the matrix elements are grouped in the hierarchical blocks and have
the binomial distributions $\{q_1,q_2,...,q_{\Gamma}\}$ as it is defined in \eq{eq:1}. Finally we arrive
at the following expression for the degree distribution ${\cal P}(m)$ (in the sake of simplicity we have
denoted $q_{\gamma}(t_{\gamma}^{(1)})\equiv q_{\gamma}(t_{\gamma})$):
\be
{\cal P}(m) = \sum_{\{t_1...t_{\Gamma}\}}\left[\prod_{\gamma=1}^{\Gamma} q_{\gamma}(t_{\gamma}) \right]
\Delta\left(\sum_{\gamma=0}^{\Gamma}p^{\gamma}t_{\gamma+1}-m\right)
\label{2:vdd1}
\ee
where the binomial distributions $q_{\gamma}(t_{\gamma})$ have the form
\be
q_{\gamma}(t_{\gamma})= p^{-\mu \gamma} \delta_{q_{\gamma}(t_{\gamma}),1} + (1-p^{-\mu
\gamma})\delta_{q_{\gamma}(t_{\gamma}),0}
\label{2:vdd2}
\ee
and $\Delta(...)$ is the Kronecker symbol:
\be
\Delta(x) = \frac{1}{2\pi i} \oint dz\; z^{x-1} =
\begin{cases} 1 & \mbox{if $x=0$} \\ 0 & \mbox{if $x\neq 0$} \end{cases}
\label{2:vdd3}
\ee
Substituting \eq{2:vdd2} and \eq{2:vdd3} in \eq{2:vdd1} after elementary transformations we get
\be {\cal
P}(m) = \frac{1}{2\pi i} \oint dz\; z^{-(m+1)}\prod_{\gamma=1}^{\Gamma}W(z,\gamma)
\label{2:vdd4}
\ee
where
\be
W(z,\gamma)=p^{-\mu \gamma} z^{p^{\gamma}}+ 1-p^{-\mu \gamma}
\label{2:vdd5}
\ee
For not too large values of $\Gamma$ the  distribution ${\cal P}(m)$ can be analyzed numerically. Using
the fact that the function $W(z,\gamma)$ is a polynomial of $z$, let us represent ${\cal P}(m)$ as:
\be
{\cal P}(m)=\frac{1}{(m+1)!}\; \left. \frac{d^{m+1}\left[\prod_{\gamma=1}^{\Gamma}
W(z,\gamma)\right]}{dz^{m+1}}\right|_{z=0}
\label{2:vdd6}
\ee
In \fig{2:vdd_fig}a we have depicted the family of curves ${\cal P}(m)$ for $\Gamma=16$ and $\mu=0.1;
1.0$. For comparison, in \fig{2:vdd_fig}c we have plotted the distribution ${\cal P}(m)$ for $\mu=1.0$,
$\Gamma=16$ as well as the binomial distribution ${\cal P}_{\rm ER}(m)=C_N^m\,q^m (1-q)^{N-m}$ for the
standard Erd\"os--R\'enyi graph with the number of vertices $N=2^{\Gamma}=2^{16}$ and for two values
$q=0.1; 0.5$. One can see on \fig{2:vdd_fig}c  how wider is the distribution ${\cal P}(m)$ with respect of
the corresponding distribution ${\cal P}_{\rm ER}(m)$ for ER--graphs.

The fractal structure of the distribution ${\cal P}(m)$ for random hierarchical networks demonstrated in
\fig{2:vdd_fig}a,b, apparently is deeply linked to the invariant multifractal measures appearing in
chaotic Hamiltonian systems in connection with the problems of ``number--theoretical chaos'' -- see, for
example, \cite{gutzw}. Actually, the condition $\Delta\left( \sum_{\gamma=1}^{\Gamma}p^{\gamma}t_{\gamma}-
m\right)$ in equation \eq{2:vdd1} for $\Gamma\to \infty$ is nothing else as the binary expansion of the
number $m$: $m=t_1\, 2^0+ t_2\, 2^1 + t_3\, 2^2 + ... + t_{\gamma+1}\, 2^{\gamma}+ ... $, where the
coefficients $t_{\gamma}$ take values 1 or 0 with corresponding probabilities $q_{\gamma}(t_{\gamma})$
defined in \eq{2:vdd2}. Let us note that the similar expansion of the form $\sum_{k=1}^{\infty}
\varepsilon_k u^{-k}$, where $u>1$ and $\varepsilon_k=\pm 1$ (with equal probabilities $\frac{1}{2}$
independent of $k$) is known in the literature as the ``singular Erd\"os measure''
\cite{erdos_sing,solomyak,vershik}. The observed fractal structure of the distribution function ${\cal
P}(m)$ appears due to the effects of the incommensurability of number--theoretic origin: some binary
expansions (of the number $m$) with random coefficients have relatively high probabilities to appear in
the adjacency matrix, while other binary expansions have much less possibilities.

\section{Conclusion}

First of all, let us note that our consideration of spectral properties of random hierarchical graphs is
far from being complete and many other properties are of interest (for example, eigenvectors, inverse
participation ratio, etc.---see, for instance \cite{goh}). However even in this preliminary investigation
we would like to emphasize the crucial difference between random hierarchical and random Erd\"os--R\'enyi
graphs.

Roughly speaking, there are two  generic ways of the scale--free network construction. The similarities
and differences of these two approaches we would like to emphasize below again.

\begin{itemize}

\item The first way, widely discussed in the literature, deals basically with the ``preferential
attachment'' procedure, where the network is raised by the essentially non-Markovian (in the increments)
evolution process with unlimited memory. The evolutionary process of such kind can be viewed, to some
extent, as a ``generalized Brownian motion''. The realization of such a procedure demands the monitoring
of the whole network structure on each step, because the appearance of new links depends on the current
degree of graph vertices. From this point of view the sequential construction of the corresponding network
can be tentatively denoted as a ``nonlinear evolution''.

\item The second way, discussed in the present work, exploits essentially different mechanism of the
scale--free network formation. The hierarchical organization of {\em probabilities} of links in the
topological network constitutes the basic idea of our construction. Specifically, we construct the
networks with the scale--free eigenvalues distribution of the adjacency matrices, where the last
originates from an appropriate randomization of the standard Parisi matrix -- one of the key objects in
the theory of spin glasses. Our method allows one to build a hierarchical network in a memoryless locally
uniform way. The application of the elements of the $p$--adic analysis \cite{avetisov1} permit us to
analyze the basic spectral properties of ensembles of randomized Parisi--type adjacency matrices.

\end{itemize}

To end up, let us emphasize that there are two important features of hierarchical networks constructed in
our paper. First of all, any sub-graph belonging to particular hierarchy level, $\gamma$, is just a random
Erd\"os--R\'enyi graph because the formation of clusters of bonds on each hierarchy level is entirely
uncorrelated. Secondly, the random sub-graphs, associated with different hierarchy levels of the network,
can be different, so the network in a whole can be essentially nonuniform. Nevertheless as we have seen,
the ``mipmapping''  construction with different sets of parameters (i.e. the hierarchical embeddings of
sub-graphs corresponding to the different hierarchical levels) leads under some conditions to the
scale--free behavior in the spectral density of adjacency matrix and to the poly--scale (and even to the
fractal) behavior in various topological characteristics of the graph. This observation is rather
unexpected since in our case the scale--free behavior is reached by essentially Markovian and memoryless
procedure.

In physics the random graphs of such a hierarchical genesis can be encountered among the scale--free
networks whose natural origin are associated with low--correlated random events carried out under short
evolutionary memory. In particular, the networks of hierarchical genesis may by interesting for the
prebiology or the earliest biology.

\begin{acknowledgments}

We are grateful to K. Bashevoy and A.Bikulov for illuminating numerical simulations and to O. Bohigas and
Y. Fyodorov for helpful discussions. S.K.N. is indebted to Z. Toroczkai for stimulating discussion of the
topic on initial stages of the work. This work is partially supported by the RFBR grant No. 07-02-00612a.

\end{acknowledgments}

\end{document}